# Counts and Colours of Faint Stars in 5 Fields Near the North Galactic Pole

L. Infante[*,1]

Grupo de Astrofísica, P. Universidad Católica de Chile, Casilla 104, Santiago 22, Chile,
Internet: linfante@astro.puc.cl

**Abstract.** Faint star number counts in the photographic $J$ band and $(B-V)$ colour distributions are presented for a 1.08 deg$^2$ field near the North Galactic Pole. Due to the excellent star/galaxy discrimination we count stars as faint as $J = 23$. We compare the number counts and colour distributions in 5 adjacent fields near SA57. The number counts and colour distributions are in very good agreement with previous data. However, we find that the large field-to-field scatter in the colour distributions, which we argue is real, might prevent us setting strong limits on Galactic structure. A simple two component standard model, Bahcall and Soneira (1984), fits the number counts reasonably well at the bright $J < 21$, but fails notably at the faint end, even if a third component is added, as in Reid and Majewski (1993). The standard models are in good agreement with both the number counts and colour distribution at $20 < V < 21$. Although the standard models bimodal shape of the colour distribution compares well with the data at $21 < V < 21.5$, the number counts of stars are underestimated.

**Key words:** Galaxy: structure – Galaxy: stellar content – Stars: statistics

## 1. Introduction

The most straightforward method to probe Galactic structure is to count stars. Kapteyn and van Rhijn (1920) pioneered this method in the early 1900s. The great advances in telescope instrumentation and computer facilities allowed in the early 1980s major achievements in this field. On the one hand, faint star number counts and colour data were possible from scans of 4 metre class telescope prime focus plates (e.g. Kron (1980), Jarvis and Tyson (1981), Reid and Gilmore (1982), Koo and Kron (1983) and Infante (1986)). And, on the other hand, star number counts and colour distribution models of the Galaxy became available: First, the more simple two component (disk and spheroid) models such as Bahcall and Soneira (1980, 1981a,b 1984, hereafter B&S) and Pritchet (1983) and then, the three component models of Gilmore (1981, 1984) and Gilmore and Reid (1983) with an added "thick disk" component. (See Bahcall (1986) for a review.)

More recently, Reid and Majewski (1993, hereafter R&M), in a very important paper, review the status of faint stars as probes of Galactic structure. They find that although the standard models are a good representation of the Galaxy at bright $V < 20$, they do not fit the observations at fainter magnitudes. Their data are from the Majewski (1992) SA57 field. They argue for the necessity of better and fainter photometry, especially for red bandpasses. In this paper, we compare the number counts and colour distributions in 5 adjacent fields near SA57. We ask whether it is possible to set constraints on Galactic structure given the large field-to-field scatter that we find in the colour distributions.

In order to probe the halo component and to test the existence of a "thick disk", faint star data, $J < 23$, are required. Although shallow surveys based on Schmidt plates (e.g. Gilmore and Reid, 1983, Gilmore, Reid and Hewett, 1985) cover large areas they only reach magnitude 19 in $V$. At $V \approx 21$ the colour distribution of stars is bimodal; the blue peak at $(B-V) \sim 0.5$ corresponds to halo subdwarf stars, whereas the red peak at $(B-V) \sim 1.5$ is mainly due to disk stars. If an intermediate population (e.g. thick disk) existed in the Galaxy, then it should appear filling the gap in the colour distribution. However, background galaxies would severely contaminate this gap if they are not extracted appropriately. The main reasons for this are, (i) galaxies at the Galactic poles outnumber stars by a factor of 2.8 at $J = 22$ (Infante and Pritchet, 1992); and (ii)

Send offprint requests to: L. Infante
* Visiting astronomer, Canada-France-Hawaii telescope, which is operated by the National Research Council of Canada, le Centre National de Recherche Scientifique, and the University of Hawaii.

the mean *(B-V)* colour of faint galaxies at this magnitude is $\sim 1.1 \pm 0.4$ (Infante and Pritchet, 1992). Therefore, in order to discriminate the "thick disk" component of our Galaxy a good star galaxy separation in the magnitude range $20 < V < 22$ is absolutely required. In previous work at these faint limits stars are separated from galaxies at $J < 21.5$ (Infante, 1986, Reid and Mejewski, 1993), whereas in this paper the limit is $J \sim 22.5$.

This paper is part of a series of papers reporting on the results of our faint object survey. The catalogue (Infante and Pritchet, 1992) contains faint stars and galaxies in an area of 2.2 deg$^2$ (with angular extent $\sim 2°\!.5 \times 2°\!.5$) near the North Galactic Pole. This catalogue, which was obtained from photographic plates taken in two bandpasses (*J* and *F*) with the Canada-France-Hawaii Telescope, is larger in solid angle coverage than any previous survey work at faint magnitudes ($J_{lim} = 24.5, F_{lim} = 23.5$); our observations are complementary to the "pencil beam" surveys obtained with CCDs at still fainter magnitudes. Due to excellent seeing of our plates, $\leq 1''$, we can discriminate stars and galaxies at fainter magnitudes than any previous survey of this type.

In this paper we investigate faint star number counts and colour distributions in 5 fields near the North Galactic Pole. The observations and reductions are presented §2. In §3 we present source number counts and colours for stars in our catalog; these results are discussed in §4.

## 2. Observations and Reductions

The data used in this paper are derived from the CFHT North Galactic Pole Survey (Infante and Pritchet 1992). In 1987 April the Canada-France-Hawaii 3.6m telescope was used to obtain 9 plates in *J* and *F* of 5 adjoining fields near the North Galactic Pole. The emulsion/filter combinations were (*IIIaJ+GG385* [=Kron (1980) *J* ] and *II-IaF+GG495* [=Kron (1980) *F* ]. Five good ($FWHM \leq 1''$) green *J* plates and four red *F* plates were exposed using the prime focus direct camera. The average scale of the prime focus plane is 13.64 *arcsec/mm* giving an unvignetted field of 0.84 *deg*$^2$.

The 5 fields were arranged in a "checkerboard" configuration (total area 2.468 deg$^2$); the seeing for all observations was $< 1''$ FWHM. The plates were scanned on the Cambridge APM (Kibblewhite et al. , 1984), and total magnitudes, image classification parameters, and astrometric positions were derived. Photometric zero points were obtained using CCD observations of a number of fields on each plate. A number of external checks verify that the magnitude scale is linear, and that zero point variations appear to be less than $\pm 0.05$ mag over the fields. The resultant catalog of approximately 40,000 galaxies and 6563 stars is estimated to be $> 90\%$ complete for galaxies to $J = 24$ and stars to $J = 23$.

Star/Galaxy classification, and further galaxy classification, were performed using the standard APM methods. Pairs of parameters derived from the APM moments were plotted. These include area vs. magnitude, mean size vs. magnitude, size vs. peak surface brightness, and magnitude vs. area/size. A probability that an object is a star or a galaxy is assigned by manually defining the class boundary. The excellent image quality of these plates allowed us to classify stars and galaxies reliably up to $J \sim 23$ and $F \sim 21.8$ where galaxies outnumber stars by a factor larger than 6 (see Table 2). The limiting *J* and *F* magnitudes for each of the fields are shown in Table 1. The reader is referred to Infante and Pritchet (1992) for further details.

Photographic emulsions do not respond linearly to incident light. Care was taken in the linearization procedure of the photographic emulsion response. Briefly, it is assumed that star profiles differ from one another only by an intensity scale factor (*e.g.* Kormendy, 1973). A look-up table is iteratively constructed in order to transform the photographic density profiles so that they are self-similar. Every image detected on the plate is used to derive a calibration curve which transforms the measured isophotal magnitudes to linearized total ($\geq 99\%$ of the image light) magnitudes . Therefore, in order to transform to a standard system requires only a zero point, which can be obtained with fairly bright stars at the bright end of the calibration curve (Bunclark and Irwin, 1983) .

The magnitude errors range from $\pm 0.3$ mag. at $J = 24$ to less than $\pm 0.1$ mag. at $J = 20$, and from $\pm 0.3$ mag. at $F = 23$ to less than $\pm 0.1$ mag. at $F = 19$. The *rms* uncertainty in the zero points is 0.022 for $J$ and 0.038 for F. The internal photometric errors in *(J-F)* are about $\pm 0.15$ at $(J+F)/2 = 21.5$ $((J+F)/2 \approx V)$.

## 3. Results

### 3.1. Number Counts

Stars were selected from the CFHT North Galactic Pole Faint object catalogue. Number counts of stars were performed as a function of magnitude on each of our 5 NGP fields. The effective areas are given in Table 1. Differential and integral star number counts in the photographic *J* bandpass are presented in Figure 1. In order to minimize sensitivity gradients across the plates and vignetting we performed star number counts only on the central 27.5.5 arcmin$^2$ area of each plate. The total area from which stars were counted is 1.0766 deg$^2$. The data are given in Table 2. The error bars in the Figure 1 represent $\sqrt{N}$ statistics. Also in Table 2 the galaxy number counts from Infante and Pritchet (1992) are shown for comparison.

To better estimate the field-to-field counting errors we have plotted in Figure 2 the star number counts for each of our 5 fields. Apart from the faint end of the number counts in field 4, where the star galaxy classifiaction boundaries rejected a higher proportion of stars, the poisson error bars in Figure 1 are consistent with the field to field scatter. It

Table 1. Effective areas an Limiting magnitudes and mean colours.

| FIELD # | AREA J deg$^2$ | Overlap* deg$^2$ | $J_{lim}$ mag. | $F_{lim}$ mag. | $<B-V>$** mag. |
|---|---|---|---|---|---|
| 1 | 0.2162 | 0.1533 | 22.3 | 21.7 | 1.12 |
| 2 | 0.2082 | – | 23.1 | – | – |
| 3 | 0.2168 | 0.2100 | 22.0 | 21.8 | 1.10 |
| 4 | 0.2165 | 0.2099 | 22.5 | 21.0 | 1.08 |
| 5 | 0.2189 | 0.2100 | 22.6 | 20.7 | 1.06 |

*Total area overlapped in the $J$ and $F$ plates
**Mean Colours are given in the magnitude range $20 \leq V \leq 21.5$. See section 3.2 for the transformation from (J-F) to (B-V).

Table 2. $J$ magnitude star number counts in the CFHT faint object NGP survey

| $J_1$ mag | $J_2$ mag | $A_{stars}(J)$ N/mag/deg$^2$ | $N_{stars}(19<J<J_2)$ deg$^2$ | $A_{galaxies}(J)$ N/mag/deg$^2$ | $N_{galaxies}(19<J<J_2)$ deg$^2$ |
|---|---|---|---|---|---|
| 19.0 | 19.5 | 246.98 | 123.49 | 185.05 | 92.53 |
| 19.5 | 20.0 | 328.69 | 287.84 | 308.41 | 246.73 |
| 20.0 | 20.5 | 479.11 | 527.39 | 500.00 | 496.73 |
| 20.5 | 21.0 | 492.11 | 773.45 | 857.01 | 925.23 |
| 21.0 | 21.5 | 622.10 | 1084.49 | 1337.38 | 1593.93 |
| 21.5 | 22.0 | 722.38 | 1445.68 | 1850.47 | 2519.16 |
| 22.0 | 22.5 | 893.22 | 1892.29 | 3085.05 | 4061.68 |
| 22.5 | 23.0 | 798.51 | 2291.55 | 5134.58 | 6628.97 |

is also clear from this figure that the completeness limit is $J \approx 23$.

We notice in Figure 2 a bump in the $J$ star number counts between $J=19.8$ and $J=20.9$ which correspond to field # 2. The point at $J=20.25$ in Figure 2 is dominated by the number counts in area #2. An eye inspection of the plates shows that all these objects are real unresolved starlike images. Unfortunately, we do not possess $F$ data for this field to check their colours.

3.2. Colour Distributions

Figures 3 and 4 show the transformed (eq. 1) (B-V) colour distribution of stars in our 4 NGP fields. The mean colours in each field are given in Table 1. Stars were counted in colour bins of 0.2 magnitudes (much larger than photometric and zero point uncertainties) and plotted as the number density per deg$^2$. We chose to present these distributions in two magnitude ranges, 20< V <21 and 21< V <21.5. Below these limits the photometry of the brightest stars is poor due to saturation and above this limit the number counts become increasingly incomplete. We note that the external errors, i.e. field to field variations are much larger than internal poisson errors.

Both Figures 3 and 4 show colour distributions which are clearly bimodal. This is a result of different density gradients in the disk and halo components, which imply differences in their scale heights. At a fixed apparent magnitude we are sampling stars in the halo which have a brighter absolute magnitude than the disk. For instance, the blue peak ($<B-V> \approx 0.5$) at $V \approx 21$ corresponds mainly to halo stars whose absolute magnitudes on average, $\sim$ +4 to +5, correspond to main-sequence turnoff stars. The effect reaches a maximum at $V \sim 21$ at the Galactic poles.

In order to compare the star count colour distributions of our NGP catalogue to the standard models in B&S and R&M we transformed our $J$ and (J-F) to V and (B-V) using the following transformation (Infante and Pritchet 1992):

$$B - V = 0.94(J - F) \quad (1)$$
$$V = J - 0.68(B - V), \quad (2)$$

which are very close to Kron (1980) transformations.

4. Discussion

Star number counts in $J$ are in very good agreement with previous work, in particular with Majewski (1992). Both data sets are for the NGP. The shallower SGP data of Infante (1986) are in reasonable agreement with the present data. The data in the present work are statistically more significant than any previous set of data: (i) We cover a much larger area, $1.08 deg^2$, as compared to $0.55 deg^2$ in Infante (1986) and $0.29 deg^2$ in Majewski (1992). And (ii) we are able to separate stars from galaxies to a much fainter limit, $J < 22.5$, in contrast to $J < 21.5$. Although the extragalactic contamination of unresolved objects (QSOs and compact, narrow emission line galaxies) is rather high at faint magnitudes no attempt is made to extract them. At the Galactic Poles $\sim 25\%$ of the faint ($V > 20.5$) blue ($(B-V) < 0.6$) starlike objects are extragalactic (Reid and Mejewski, 1993, and Kron et al., 1991).

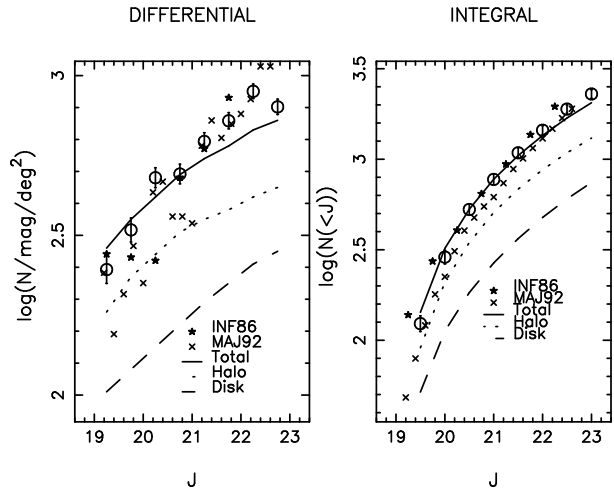

**Fig. 1.** Differential (left panel) and integral (right panel) star counts as a function of $J$ magnitude for the NGP. The number counts are given per deg$^2$ for 1 mag intervals. Error bars correspond to $\sqrt{N}$ statistics. The solid line is the number counts from B&S standard model. The dashed and dotted lines are the disk and spheroid components of the model respectively. INF86 are SGP data points in Infante (1986) and MAJ92 are data points for SA57 in Majewski 1992.

In principle, the bimodal colour distribution provides a good means to discriminate halo stars from disk stars by their colours. Nevertheless, the large external field-to-field errors prevent setting strong constraints on the various Galactic components, e.g. disk, extended/thick disk and halo. With the current data, one cannot distinguish a two component model, such as the traditional B&S standard model, from models which include a third extended/thick disk components as in R&M.

Figures 3 and 4 show a large field-to-field scatter which may be due to a variety of reasons:

- Variations in seeing from plate to plate may affect the completeness limits due to variable quality of the star/galaxy separation. However, our colour distributions are for magnitude limits brighter than the lowest limit in Table 1. This seeing effect might be more critical in other works at faint magnitudes.
- Magnitude zero point differences from gradients across the plates could in principle produce differences in the number of stars in each colour bin. However, this possibility was investigated in paper I; based on the galaxy distribution we conclude that there is a negligible density gradient across the plates.
- Field to field magnitude zero point differences could also produce scatter in the number counts. However, the plate-to-plate zero points were tied together by CCD photometric sequences observed at different locations for each field.
- Contamination by QSOs would affect the blue bins more than the red ones. (See a discussion about this effect in R&M¿)

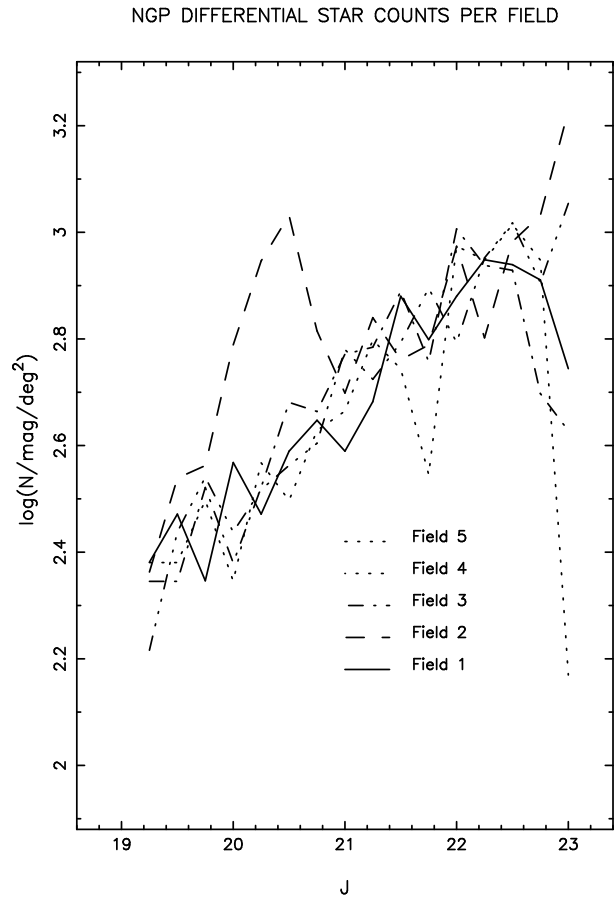

**Fig. 2.** Differential star counts as a function of $J$ magnitude for each of the 5 NGP fields. The number counts are given per deg$^2$ for 1 mag intervals.

Intrinsic differences in the number of stars in each bin in each field is the more likely possibility. For instance, this as was noted in §3.1, in the magnitude range $J = 19.8$ and 20.9, there are 2.5 times more stars in field #2 than in the other four fields.

### 4.1. Models

We now turn to a comparison with standard models. We have built a two component model using the B&S Export code (see Bahcall 1986 for model definitions), and have taken the standard three component model from R&M . Properties used to compute the models are specified in Table 3. Due to the negligible absorption at the Galactic Poles a zero absorption coefficient was adopted. The B&S standard model is shown as a solid line in Figure 1 and as the connected squares (□) in Figures 3 and 4. The three component R&M standard model is shown in Figures 3 and 4 as connected diamonds (◇).

At $20 < V < 21$ the two models (B&S and R&M) are in good agreement with both the number counts and colour distributions. Overall, at $21 < V < 21.5$, there are 1.3 times more stars than what the standard model predicts. Qualitatively, the B&S predicts a higher red than blue peak in better agreement with the data, in contrast to the R&S prediction. However, these results have to be taken with caution since the field-to-field variations could dominate over these effects.

Is it possible to say anything about the "thick disk" with the present data? The current discussion of the existence of the thick disk cannot be resolved with the present data. This is proven by the large scatter among fields in the number counts at each bin. More data at faint magnitudes ($J > 21$) are needed. High spatial resolution images are required to discriminate stars from galaxies at faint limits. Spectrophotometry is necessary to identify the QSOs.

### 5. Summary and Conclusions

In this paper we present star number counts and colour distributions obtained from the CFHT North Galactic Pole survey. Five fields were observed with the prime focus and nine plates were obtained in two filter bandpasses, $J$ and $F$. The plates were scanned and only the central unvignetted areas were analyzed. The total area from which stars were studied is $\sim 1.08 deg^2$. Special care was taken to determine the magnitude level at which stars are separated from galaxies. This point varies from $J = 22$ (Field #3) to $J = 23.1$ (Field #2) and from $F = 20.7$ (Field #5) to $F = 21.8$ (Field #3).

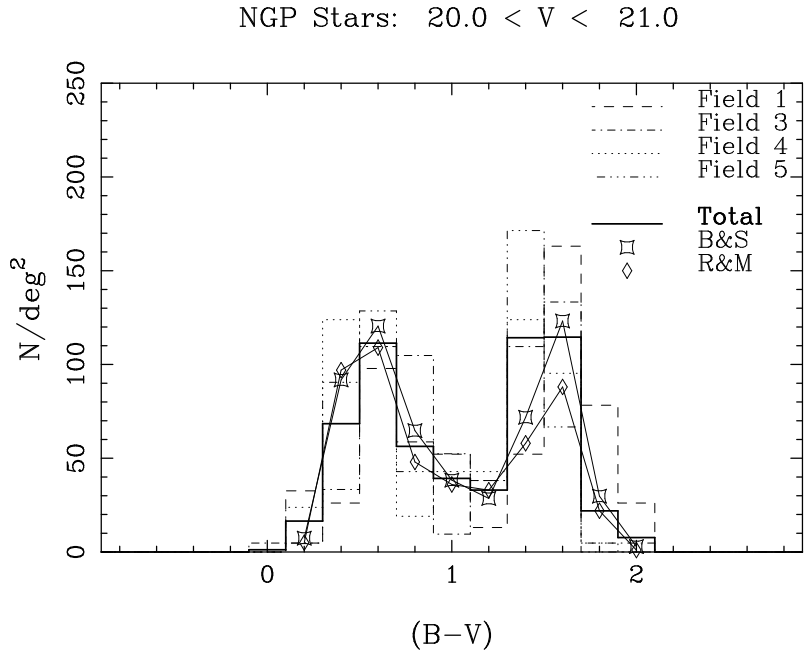

**Fig. 3.** (B-V) colour distribution of stars in the NGP Fields. The number of stars per deg$^2$ in the magnitude range $20 < V < 21$ is plotted. The solid line is the colour distribution of the combined 4 fields. We also show the number counts of B&S and R&M standard models.

**Table 3.** Bahcall-Soneira and Reid-Majewski Standard Models

| COMPONENT | CHARACTERISTIC | B&S[1] | R&M[2] |
|---|---|---|---|
| DISK | Density Law | Exponential | Exponential |
| | Scale Length | 3500 pc | 4000 pc |
| | Scale height (m.s. stars) | 325 pc | 325 pc |
| | Scale height (giants) | 250 pc | 250 pc |
| | Luminosity Function | Wielen (1974) | Wielen (1974) |
| | Colour Magnitude Diagram | M67 | Bessel (1990) |
| THICK DISK | Density Law | — — | Exponential |
| | Local Normalization | — — | 0.02 |
| | Scale height | — — | 1200 pc |
| | Scale length | — — | 4000 pc |
| | Luminosity Function | — — | Wielen (1974) |
| | Colour Magnitude Diagram | — — | 47 Tuc |
| HALO | Density Law | de Vaucouleurs | de Vaucouleurs |
| | $\frac{b}{a}$ | 0.8 | 0.85 |
| | Local Normalization | 0.002 | 0.0015 |
| | Effective Radius | 2.7 kpc | 2.7 kpc |
| | Colour Magnitude Diagram | M13 | M3 |

[1] See Bahcall (1986) for model definitions.
[2] See R&M for model definitions.

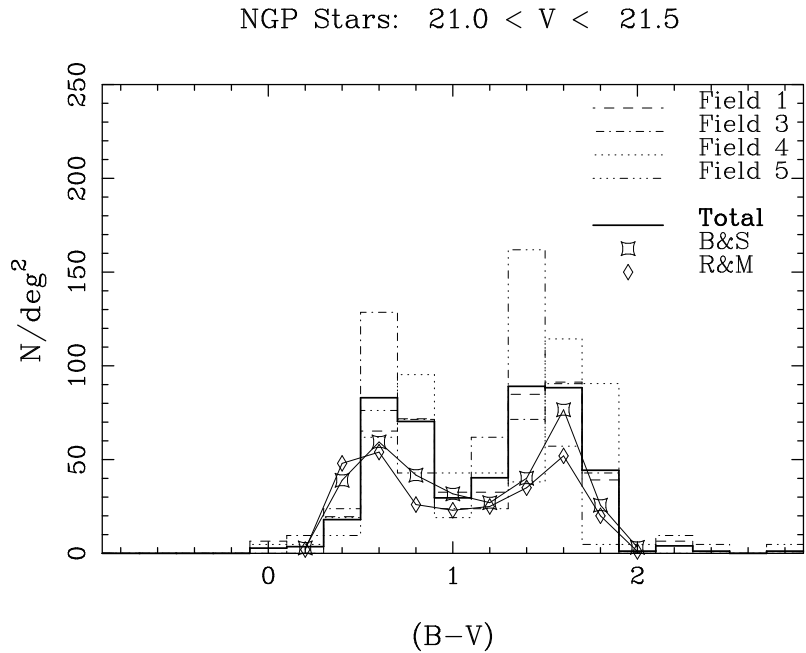

**Fig. 4.** (B-V) colour distribution of stars in the NGP Fields. The number of stars per deg$^2$ in the magnitude range $21 < V < 21.5$ is plotted. The solid line is the colour distribution of the combined 4 fields. We also show the number counts of B&S and R&M standard models.

Number counts of stars to a limiting magnitude 22.5 were performed in the $J$ bandpass. At the faint end they compare very well with the number counts of Majewski (1992) and of Infante (1987). At the bright end the number counts also compare very well with the published number counts except for field # 2 where there is an overdensity of stars at $J \approx 20.5$. Due to our better number statistics, which is consistent with the field-to-field scatter, our star number counts are more reliable than any previous work at this faint limit. A simple two component standard model, B&S, fit the number counts reasonably well at the bright $J < 21$, but fails notably at the faint end, even if a third component is added, as in R&M. The excess of stars at $J > 21$ with respect to the models may be in part due to poor star/galaxy classification and contamination by unresolved extragalactic objects.

*Acknowledgements.* We are grateful to Chris Pritchet and Wolfgang Gieren for helpful discussions. This work was supported by Proyecto FONDECYT 93/0570